\documentclass[5p,twocolumn]{elsarticle}
\usepackage{tabularx}
\usepackage{siunitx}
\usepackage{miller}
\usepackage{graphicx}
\usepackage[modulo,switch]{lineno}
\usepackage{soul}
\usepackage{tikz}
\usepackage{amssymb}
\usetikzlibrary{decorations.pathreplacing}
\usetikzlibrary{positioning}

\begin{document}
\title{Stability of Mn$_2$Ru$_x$Ga-based Multilayer Stacks}

\author{Gwena\"el Atcheson\corref{cor1}}
\cortext[cor1]{Corresponding author}
    \ead{atchesog@tcd.ie}

\author{Katarzyna Siewierska\corref{}}
\author{J. \,M. \,D. \,Coey\corref{}}
\author{Karsten Rode\corref{}}
\author{Plamen Stamenov\corref{}}

\address{CRANN, AMBER and School of Physics, Trinity College, Dublin 2, Ireland}
\date{\today}

\nolinenumbers

\begin{abstract}
Perpendicular heterostructures based on a ferrimagnetic Mn$_2$Ru$_x$Ga (MRG) layer and a ferromagnetic Co/Pt multilayer were examined to understand the effects of different spacer layers (V, Mo, Hf, HfO$_x$ and TiN) on the interfaces with the magnetic electrodes, after annealing at \SI{350}{\degreeCelsius}. Loss of perpendicular anisotropy in MRG is strongly correlated with a reduction in the substrate-induced tetragonality due to relaxation of the crystal structure. \hl{In the absence of diffusion, strain and chemical ordering within MRG are correlated. The limited solubility of both Hf and Mo in MRG is a source of additional valence electrons, which results in an increase in compensation temperature $T_{\text{comp}}$. This also stabilises perpendicular anisotropy, compensating for changes in strain and defect density.} The reduction in squareness of the MRG hysteresis loop measured by anomalous Hall effect is \SI{<10}{\percent}, making it useful in active devices. Furthermore, a \hl{CoPt$_3$ phase with} \hkl(220) texture in the perpendicular Co/Pt free layer promoted by a Mo spacer layer is the only one that retains its perpendicular anisotropy on annealing.
\end{abstract}
\maketitle

\section{Introduction}\label{sec:intro}
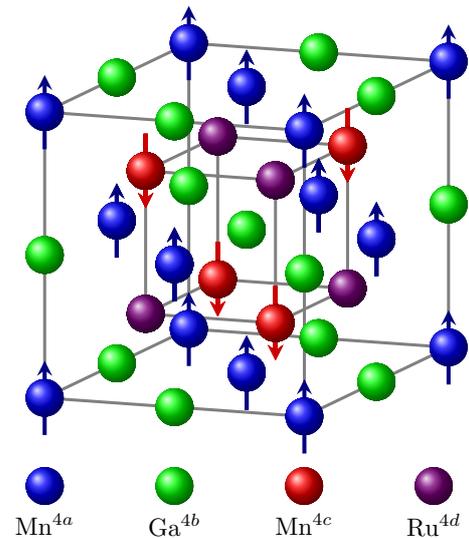
\begin{figure}[h]
    \centering
    \resizebox{0.7\linewidth}{!}{\newcommand{\myPgfX}{}%
\newcommand{\myPgfY}{}%
\newcommand{\myPgfZ}{}%
  \begin{tikzpicture}[scale = 4, z = -2.50mm, rotate around y = -15, 
    atom/.style = {circle, shade, shading = ball, inner sep = 0, minimum size = 15.0} ]
    \draw [help lines,very thick] (0,0,0) -- (0,0,1) -- (0,1,1) -- (0,1,0) -- (0,0,0) -- (1,0,0) -- (1,0,1) -- (1,1,1) -- (1,1,0) -- (1,0,0) (1,1,0) -- (0,1,0) (0,1,1) -- (1,1,1) (0,0,1) -- (1,0,1);
    \draw [help lines,very thick] (0.25,0.25,0.25) -- (0.25,0.25,0.75) -- (0.75,0.25,0.75) -- (0.75,0.25,0.25) -- cycle -- (0.25,0.75,0.25) -- (0.75,0.75,0.25) -- (0.75,0.75,0.75) -- (0.25,0.75,0.75) -- (0.25,0.75,0.25) (0.25,0.75,0.75) -- (0.25,0.25,0.75) (0.75,0.75,0.75) --(0.75,0.25,0.75) (0.75,0.75,0.25) -- (0.75,0.25,0.25);
    \foreach \A/\B/\C in {0/0/0,1/0/0/,0/1/0,0/0/1,1/1/0,0/1/1,1/1/1,1/0/1}
      \foreach \x/\y/\z in {0/0/0,0/0.5/0.5,0.5/0/0.5,0.5/0.5/0}
        \pgfmathsetmacro{\myPgfX}{{(\A+\x > 1) ? frac(\A+\x) : \A+\x}}%
        \pgfmathsetmacro{\myPgfY}{{(\B+\y > 1) ? frac(\B+\y) : \B+\y}}%
        \pgfmathsetmacro{\myPgfZ}{{(\C+\z > 1) ? frac(\C+\z) : \C+\z}}%
          \draw [->,>=stealth,ultra thick,blue!50!black] (\myPgfX,\myPgfY-0.13,\myPgfZ) -- +(0,0.26,0);
    \foreach \A/\B/\C in {0/0/0,1/0/0/,0/1/0,0/0/1,1/1/0,0/1/1,1/1/1,1/0/1}
      \foreach \x/\y/\z in {0/0/0,0/0.5/0.5,0.5/0/0.5,0.5/0.5/0}
        \pgfmathsetmacro{\myPgfX}{{(\A+\x+0.25 > 1) ? frac(\A+\x+0.25) : \A+\x+0.25}}%
        \pgfmathsetmacro{\myPgfY}{{(\B+\y+0.25 > 1) ? frac(\B+\y+0.25) : \B+\y+0.25}}%
        \pgfmathsetmacro{\myPgfZ}{{(\C+\z+0.25 > 1) ? frac(\C+\z+0.25) : \C+\z+0.25}}%
	  \draw [<-,>=stealth,ultra thick,red!80!black] (\myPgfX,\myPgfY-0.13,\myPgfZ) -- +(0,0.26,0);
    \foreach \A/\B/\C in {0/0/0,1/0/0/,0/1/0,0/0/1,1/1/0,0/1/1,1/1/1,1/0/1}
      \foreach \x/\y/\z in {0/0/0,0/0.5/0.5,0.5/0/0.5,0.5/0.5/0}
        \pgfmathsetmacro{\myPgfX}{{(\A+\x > 1) ? frac(\A+\x) : \A+\x}}%
        \pgfmathsetmacro{\myPgfY}{{(\B+\y > 1) ? frac(\B+\y) : \B+\y}}%
        \pgfmathsetmacro{\myPgfZ}{{(\C+\z > 1) ? frac(\C+\z) : \C+\z}}%
	\node at (\myPgfX,\myPgfY,\myPgfZ) [atom, ball color = blue] {};
    \foreach \A/\B/\C in {0/0/0,1/0/0/,0/1/0,0/0/1,1/1/0,0/1/1,1/1/1,1/0/1}
      \foreach \x/\y/\z in {0/0/0,0/0.5/0.5,0.5/0/0.5,0.5/0.5/0}
        \pgfmathsetmacro{\myPgfX}{{(\A+\x+0.5 > 1) ? frac(\A+\x+0.5) : \A+\x+0.5}}%
        \pgfmathsetmacro{\myPgfY}{{(\B+\y+0.5 > 1) ? frac(\B+\y+0.5) : \B+\y+0.5}}%
        \pgfmathsetmacro{\myPgfZ}{{(\C+\z+0.5 > 1) ? frac(\C+\z+0.5) : \C+\z+0.5}}%
          \node at (\myPgfX,\myPgfY,\myPgfZ) [atom, ball color = green] {};
    \foreach \A/\B/\C in {0/0/0,1/0/0/,0/1/0,0/0/1,1/1/0,0/1/1,1/1/1,1/0/1}
      \foreach \x/\y/\z in {0/0/0,0/0.5/0.5,0.5/0/0.5,0.5/0.5/0}
        \pgfmathsetmacro{\myPgfX}{{(\A+\x+0.25 > 1) ? frac(\A+\x+0.25) : \A+\x+0.25}}%
        \pgfmathsetmacro{\myPgfY}{{(\B+\y+0.25 > 1) ? frac(\B+\y+0.25) : \B+\y+0.25}}%
        \pgfmathsetmacro{\myPgfZ}{{(\C+\z+0.25 > 1) ? frac(\C+\z+0.25) : \C+\z+0.25}}%
          \node at (\myPgfX,\myPgfY,\myPgfZ) [atom, ball color = red] {};
    \foreach \A/\B/\C in {0/0/0,1/0/0/,0/1/0,0/0/1,1/1/0,0/1/1,1/1/1,1/0/1}
      \foreach \x/\y/\z in {0/0/0,0/0.5/0.5,0.5/0/0.5,0.5/0.5/0}
        \pgfmathsetmacro{\myPgfX}{{(\A+\x+0.75 > 1) ? frac(\A+\x+0.75) : \A+\x+0.75}}%
        \pgfmathsetmacro{\myPgfY}{{(\B+\y+0.75 > 1) ? frac(\B+\y+0.75) : \B+\y+0.75}}%
        \pgfmathsetmacro{\myPgfZ}{{(\C+\z+0.75 > 1) ? frac(\C+\z+0.75) : \C+\z+0.75}}%
          \node at (\myPgfX,\myPgfY,\myPgfZ) [atom, ball color = red!50!blue] {};

	  \node at (-0.65+0.15,-0.55) [rotate around y = 15, atom, ball color = blue, label = {below:Mn$^{4a}$}] {};
	  \node at (-0.20+0.15,-0.55) [rotate around y = 15, atom, ball color = green, label = {below:Ga$^{4b}$}] {};
	  \node at (0.25+0.15,-0.55)  [rotate around y = 15, atom, ball color = red, label = {below:Mn$^{4c}$}] {};
	  \node at (0.70+0.15,-0.55)  [rotate around y = 15, atom, ball color = red!50!blue, label = {below:Ru$^{4d}$}] {};
  \end{tikzpicture}}
        \caption{Crystal structure of Mn$_2$Ru$_x$Ga is a distorted XA Heusler structure with two inequivalent, antiferromagnetically coupled sublattices, Mn$^{4a}$ and Mn$^{4c}$, leading to zero moment at compensation \cite{kurt2014}}
            \label{fig:crystal}
\end{figure}

Modern spintronic materials are being developed for high-speed, non-volatile data storage products such as magnetic random-access memory. An ideal material would be one which has high spin-polarisation, yet little or no magnetic moment.
Mn$_2$Ru$_x$Ga (MRG) is a ferrimagnetic inverse Heusler alloy, having an XA structure with two inequivalent antiferromagnetically aligned Mn sublattices \cite{kurt2014} and almost complete filling of the four lattice sites \cite{siewierska2021}, as shown in Figure \ref{fig:crystal}.
It is considered to be a zero-moment half-metal (ZHM) when its net moment vanishes at the compensation temperature, $T_{\text{comp}}$, and it is a potential candidate for spintronic applications owing to its immunity to magnetic fields, reasonably high Curie temperature and large perpendicular anisotropy field close to compensation.
\hl{The possibility of tunable anti-ferromagnetic oscillation modes in the {\si{\tera\hertz}} region} \cite{troncoso2019} \hl{is key for future applications in high-frequency computing and communications, and has already been observed in the similar Mn$_{3-x}$Ga materials} \cite{awari2016}. 
Perpendicular magnetic anisotropy (PMA) due to biaxial strain is observed in films grown on MgO.
The ruthenium $4d$ site occupancy allows the magnetic properties to be tuned according to the application; in particular, $T_{\text{comp}}$ increases with increasing Ru concentration and valence electron count, $n_v$, in the unit cell \cite{siewierska2021,nivetha2015}. 

Annealing of magnetoresistive devices can have varying outcomes - \hl{the crystallisation of the MgO/CoFeB bilayer results in a dramatic increase of  tunnelling magnetoresistance (TMR) ratio due to coherent tunnelling and perpendicular anisotropy in the free layer}, while the diffusion of Ta and other species into the active portion of the device can have a negative impact on functionality \cite{kurt2010,Ikeda2008,fukumoto2004,Yang2012}.
This MgO/CoFeB electrode structure is currently the industry choice for TMR applications, and annealing temperatures for crystallisation of amorphous CoFeB into a body centered cubic structure typically occurs above \SI{320}{\degreeCelsius}, depending on boron content. 

The use of thin interlayers to enhance performance of MnGa-based magnetic tunnel junctions has proved effective \cite{ma2012,kubota2011,liang2014}.
A major concern with Mn-based materials, however, is diffusion which can affect the both crystal structure as well as the quality of the spin-polariser/barrier interfaces \cite{suzuki2018,mao2017}. 
We have previously demonstrated that an ultrathin dusting layer of Ta \cite{titova2019} or Al \cite{borisov2016} can be used to mitigate the effects of diffusion between MRG and an MgO tunnel barrier. 
\hl{Using this technique, a TMR of {\SI{40}{\percent}} has been achieved at {\SI{10}{\kelvin}} and at low bias in a MRG based magnetic tunnel junction using a CoFeB free layer} \cite{borisov2016}.
Similarly, we have shown that Hf can magnetically couple MRG and CoFeB thin films, after annealing at high temperatures \cite{borisov2017}.
Hf and Al are not known for preventing diffusion, so the mechanism by which they act to maintain performance is not clear, compared to other similar diffusively mobile materials.
A previous study of MRG, annealed at various temperatures up to \SI{400}{\degreeCelsius} and capped with AlO$_x$, shows that the magnetic and crystalline properties are stable up to \SI{350}{\degreeCelsius}, beyond which the crystal structure relaxes \cite{kat_2017}.

There are several types of defects that occur in the MRG lattice - Mn-Ga ($4a/4c \leftrightarrow 4b$), Mn-Ru ($4a/4c \leftrightarrow 4d$), and Ru-Ga ($4d \leftrightarrow 4b$) antisites - as well as excess Mn and Ga on ${4d}$ sites when $x < 1$ . 
We can quickly identify Ru-based defects in X-ray crystallography due to the much stronger atomic scattering factor of Ru compared to Mn and Ga. 
Mn-Ga antisite defects have previously been identified as the source of additional electronic pressure that modifies the position of the Fermi level, increasing the moment and spin polarisation \cite{zic_2016}.
Here we correlate magnetic and transport properties with the presence of crystalline defects.

\begin{table}[t]
\caption{\hl{Spacer material selection}}
\begin{tabularx}{\linewidth}{c  X}
\hline
\textbf{Material ($y$)} & \textbf{Reason}\\\hline
TiN & Known diffusion barrier; close lattice match with MgO should allow for crystalline growth on MRG at high temperature \\
Hf & Ultrathin layers promote coupling between MRG and CoFeB\\
HfO$_x$ & Used as high-$k$ barrier (gate dielectric)\\
V & A known Al getter\\
Mo & Low affinity for alloying with Mn\\
\end{tabularx}
\label{table:spacers}
\end{table}

To investigate, we use a spin-valve structure of the form MgO//Mn$_2$Ru$_0._7$Ga(\num{30})/X$(t)$/[Co(\num{0.4})/Pt(\num{0.8})]$_8$/Co(\num{0.4})-/Pt(\num{3})
where the spacer layer, X, has a thickness, $t$, of \SIlist{1.4;2}{\nano\metre} (parenthetical values in \si{\nano\metre}). It will act as a barrier preventing diffusion, or as a source of mobile atoms. The stack uses a Co/Pt superlattice as the free layer due to its perpendicular anisotropy \hl{as deposited, independent of seed layer or annealing requirements as with a MgO/CoFeB free layer}.

The five materials selected as X in this study are listed in Table \ref{table:spacers}. 
TiN thin films are widely used as diffusion barriers and serve as a baseline comparison where we expect no diffusion.
Hf has also been chosen to serve as a comparison due to our previous experience with it \cite{borisov2017}.
Diffusion-driven defects are expected to result in a reduction of crystalline order in MRG. Any change in $c/a$ ratio will change the magneto-crystalline anisotropy in MRG, while the perpendicular anisotropy in Co/Pt multilayers is primarily caused by interfacial anisotropy driven by spin-orbit hybridisation of Co and Pt\cite{nakajima1998}. This means that a change in the interfacial roughness or strain will have an impact on the Co/Pt magnetic properties \cite{qiu_2016,nakagawa2005}.

We measured crystallographic and magnetotransport characteristics before and after annealing at \SI{350}{\degreeCelsius}, a standard  industrial practice for MgO/CoFeB-based devices. In this way we aimed to identify the influence of annealing the structure based on the interlayer material used, and its effect on the properties of both electrodes.

\section{Methods}\label{sec:method}
All films except TiN were deposited by \hl{direct current} sputtering onto \hl{$\SI{10x10}{\milli\metre}$} single crystal MgO \hkl(001) substrates in a SFI Shamrock sputter tool (base pressure \hl{$\SI{< 2.5e-6}{\pascal}$}); TiN  was deposited by \hl{radio frequency} sputtering. MRG is deposited at \SI{320}{\degreeCelsius} by co-sputtering from a  Mn$_2$Ga target and a Ru target, \hl{both {$\SI{76.2}{\milli\metre}$} diameter, at a target-substrate distance of {$\SI{100}{\milli\metre}$}. Argon gas flow was {$\SI{38}{\cubic\centi\metre\per\minute}$}}. All other materials were deposited at room temperature, except for two spacer layers of TiN grown at \SI{320}{\degreeCelsius}. HfO$_x$ was formed by natural oxidation of metallic Hf films \hl{in O$_2$ atmosphere at {$\SI{2}{\pascal}$} for {$\SI{60}{\second}$}.} Annealing is carried out \hl{under vacuum} at \SI{350}{\degreeCelsius} for 60 minutes in a perpendicular field of \SI{800}{\milli\tesla}. X-ray diffraction \hl{(XRD) and X-ray reflection (XRR) were measured} with a Panalytical X'Pert tool using Cu K$_\alpha$ radiation, and magnetotransport data was collected by the Van der Pauw method in a \SI{1}{\tesla} GMW electromagnet. Crystallographic data were analysed \hl{following the method of }\cite{Nandi1978} by fitting a Voigt function to the peaks, taking into consideration K$\alpha_2$ and instrumental broadening, characterised using an Al$_2$O$_3$ standard, NIST 1976b.

\section{Results}\label{sec:results}

\begin{figure}
\resizebox{\linewidth}{!}{\input{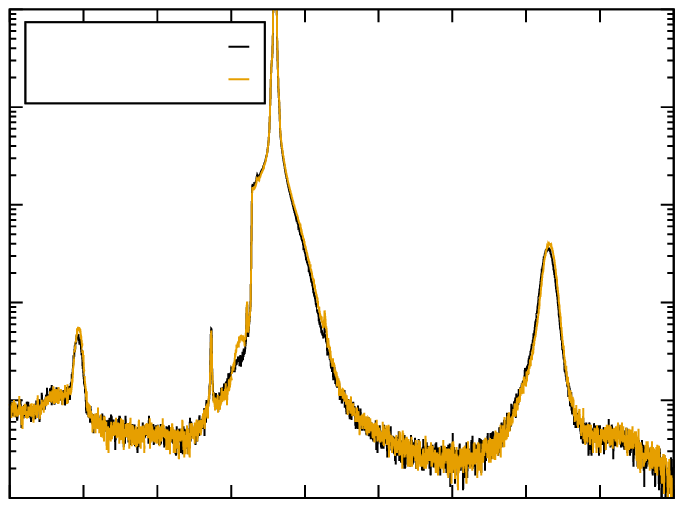}}
\caption{XRD of a spin-valve structure with a \SI{1.4}{\nano\metre} Hf spacer layer, before and after annealing. Peak labels are: $\bullet$ MgO substrate, $\bigtriangleup$ MRG, and $\circ$ CoPt$_3$}
\label{fig:Hf_XRD}
\end{figure}

\begin{figure}
\resizebox{\linewidth}{!}{\input{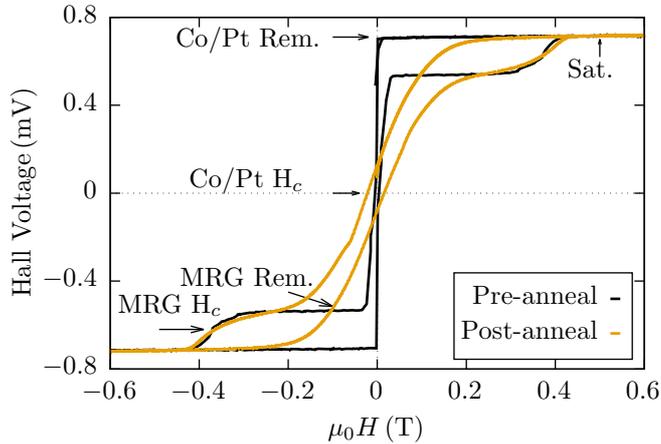}}
\caption{AHE of $\SI{1.4}{\nano\metre}$ Hf spacer stack - Co/Pt and MRG layers are able to switch independently. \hl{Remanence (Rem.), coercivity ($H_c$), and saturation (Sat.) are indicated for each layer. Remanence for MRG is given as the remanent state after the Co/Pt layer has switched, and it's coercivity is given as the point in the hysteresis halfway between the remanent and fully saturated states.} Co/Pt has lost some perpendicular anisotropy and appears partially coupled to the MRG after annealing}
\label{fig:Hf_EHE}
\end{figure}

\begin{figure}
\resizebox{\linewidth}{!}{\input{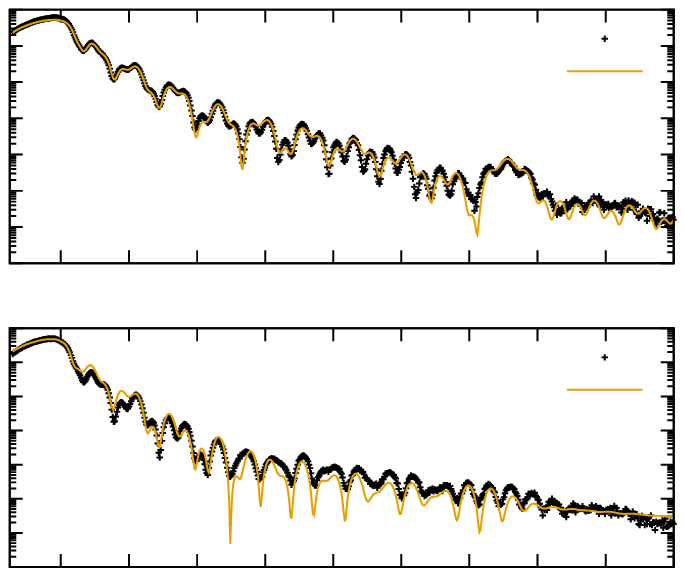}}
\caption{XRR and model fitting of a spin-valve structure with a \SI{1.4}{\nano\metre} V spacer layer, (a) before and (b) after annealing. $\ast$ indicates the Co/Pt \hkl(001) superlattice peak position for $d_{Co}+d_{Pt}=\SI{1.19}{\nano\metre}$.}
\label{fig:XRR}
\end{figure}

\begin{figure*}
\resizebox{\textwidth}{!}{\input{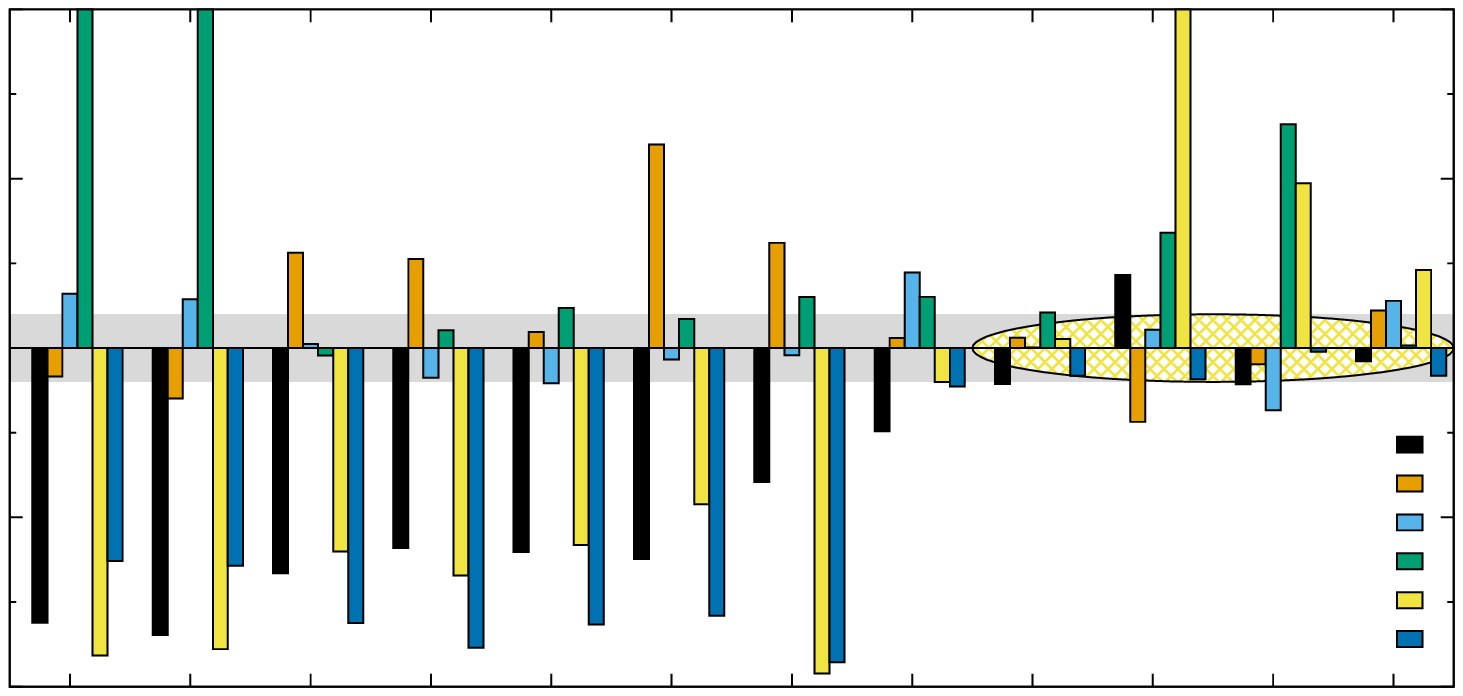}}
\caption{Relative change of tetragonal distortion ($\delta_c$), coherence length ($L_c$), strain ($\epsilon$), intensity ratio ($S$), coercivity ($H_c$), and squareness in MRG after annealing. The grey box indicates a threshold of \SI{10}{\percent} change in squareness as a metric for potential use in  applications. The yellow ellipse indicates the best two candidate spacer materials.}
\label{fig:MRG_combined}
\end{figure*}

As-deposited MRG is textured, with the crystallographic $c$-axis perpendicular to the substrate surface. 
Two diffraction peaks are observed in a symmetric $\theta\text{-}2\theta$ scan, namely the \hkl(002) and \hkl(004) (Figure \ref{fig:Hf_XRD}). In addition to the substrate and MRG peaks, we also note the Co/Pt superlattice \hkl(002) peak. 
As mentioned in section \ref{sec:intro}, the magnetic properties are strongly dependent on the crystal characteristics. We determine the out-of-plane lattice parameter $c$, crystal coherence length $L_c$ perpendicular to the film (using the Scherrer formula), \hl{the perpendicular strain $\epsilon$,} and the ratio of intensities of the two peaks $S_{\hkl(002)/\hkl(004)}$ signifying the Ru/Mn order as Ru occupying the $4a$ or $4b$ sites will increase the intensity of the \hkl(002) peak. Higher $S_{\hkl(002)/\hkl(004)}$ ratios indicates disorder.
By comparison $L_c$ \hl{and $\epsilon$} give us a metric for the long-range disorder in the film, due to grain boundaries and stacking faults, where larger grain size or less strain indicates \hl{a more homogenous film}.

MRG has a coercivity in a perpendicular magnetic field that diverges as the temperature approaches $T_{\text{comp}}$, where the anomalous Hall effect (AHE) signal changes sign \cite{nivetha2015}. The composition of MRG used here, $x = 0.7$, was chosen to give $T_{\text{comp}}=\SI{165}{\kelvin}$\hl{ }\cite{siewierska2021}\hl{,} with an as-deposited coercivity of $H_c \approx \SI{380}{\milli\tesla}$ at room temperature.

Squareness of the Hall signal, i.e. the ratio of Hall voltage at saturation to that at remanence, is used an indicator of perpendicular anisotropy. The main parameters from the AHE hysteresis are labeled in Figure \ref{fig:Hf_EHE}.
The as-deposited MRG exhibits perpendicular magnetic anisotropy (PMA) in all cases, irrespective of spacer material or thickness; magnetic properties upon annealing \textit{are} spacer material dependent. It is typically reported that perpendicular anisotropy in Co/Pt is accompanied by a \hkl(111) texture \cite{lin1991}, however we note here that all as-deposited Co/Pt films exhibited perpendicular anisotropy regardless of \hl{initial texture or phases present}.

After annealing with a $\SI{1.4}{\nano\metre}$ Hf spacer, the Co/Pt layer loses much of its perpendicular anisotropy, indicated by the loss of squareness as demonstrated in Figure \ref{fig:Hf_EHE}. 
This can be attributed to \hl{an increase in average interfacial roughness, from {\SI{0.46}{\nano\metre}} to {\SI{0.50}{\nano\metre}} in Co and from {\SI{0.53}{\nano\metre}} to {\SI{0.99}{\nano\metre}} in Pt. This} reduces the interfacial anisotropy, bringing the moment in-plane, \hl{and is caused by intermixing} as evidenced by the weak crystallisation of \hl{CoPt$_3$} with \hkl(111) texture after annealing, as seen in Figure \ref{fig:Hf_XRD}.
\hl{The Co and Pt interfaces of the free layer are well defined before annealing, as can be seen from XRR data for a hetereostructure with a {$\SI{1.4}{\nano\metre}$} V spacer layer in Figure {\ref{fig:XRR}}(a). A clear and intense Co/Pt {\hkl(001)} superlattice peak is visible which disappears upon annealing, shown in Figure {\ref{fig:XRR}}(b), a result of intermixing and alloying between the layers forming a CoPt$_3$ phase as shown in Figure {\ref{fig:Hf_XRD}}. 
XRD and roughness data for the Co/Pt multilayers is given in Table {\ref{table:copt}}. Of interest is that in each case, the average Co roughness is approximately equivalent to the desired layer thickness ({\SI{0.4}{\nano\metre}}), indicating incomplete coverage.}
The anisotropy of MRG appears to be maintained, with minimal change in coercivity or squareness.
Changes in Co/Pt properties are considered independent of changes in MRG.
As discussed in Section \ref{sec:intro}, the magneto-crystalline anisotropy \hl{of MRG is largely} dependent on the $c/a$ ratio, or more specifically the tetragonal distortion of the cubic cell - $\delta_c=(c-a)/a$.

\hl{The properties of MRG are highly dependent on the spacer layer used, and the} relative change of each parameter in MRG after annealing has been collated in Figure \ref{fig:MRG_combined}.
Below, we summarise the results for each spacer material. For reference, Table \ref{table:stats} gives a summary of the mean and standard deviation for the various properties measured in MRG before annealing. As a basic metric, we  aim to maintain squareness of magnetic hysteresis in MRG to within \SI{90}{\percent} of initial value after annealing \hl{in order to preserve functionality of active devices}.

Giant magnetoresistance was not measured in these devices as a current-perpendicular-to-plane geometry would be required to prevent shunting effects of currents through highly resistive layers, necessitating nanoscale lithographic patterning of the metallic films.

\begin{table}
\caption{\hl{Statistics for MRG properties before annealing. $a$ is {\SI{5.96}{\angstrom}}, defined by growth on the substrate}}
\begin{tabularx}{\columnwidth}{l X X}
\hline
 & \textbf{Mean} & \textbf{$\sigma$}\\\hline
$c$ & \SI{6.0390}{\angstrom} & \SI{0.0108}{\angstrom}\\
$\delta_c$ & \SI{1.326}{\percent} & \SI{0.181}{\percent}\\ 
$L_c$ & \SI{14.66}{\nano\metre} & \SI{1.93}{\nano\metre}\\
$\epsilon$ & \SI{0.511}{\percent} & \SI{0.056}{\percent}\\ 
$S_{\hkl(002)/\hkl(004)}$ & 0.0758 & 0.0056\\
$H_c$ & \SI{371.5}{\milli\tesla} & \SI{71.7}{\milli\tesla}\\
Squareness & \SI{92.1}{\percent} & \SI{7.5}{\percent}\\
\end{tabularx}
\label{table:stats}
\end{table}

\subsection{Vanadium}\label{subsec:V}
Annealing of a vanadium spacer stack results in large reductions of the MRG coercivity ($\sim \SI{-90}{\percent}$) and squareness ($\sim \SI{-75}{\percent}$), along with a significant ($\sim \SI{-80}{\percent}$) reduction in $\delta_c$. This indicates the anisotropy has a primarily in-plane contribution, though a small fraction of grains are still perpendicular.
This is accompanied by an extremely large ($\SI{> 500}{\percent}$) increase of $S$, \hl{as well as a moderate decrease of $L_c$ and increase in $\epsilon$} indicating complete reordering of the crystal structure. Additional peaks near the MRG \hkl(002) and \hkl(004) peaks with $d$-spacing of $\sim$ \SIlist{0.3058;0.1511}{\nano\metre} are present after annealing. These $d$-spacings correspond well to either RuV$\hkl(100) (\SI{0.298}{\nano\metre})$, RuV$\hkl(200) (\SI{0.148}{\nano\metre})$ or Ga$_5$V$_2 \hkl(220) (\SI{0.317}{\nano\metre})$, Ga$_5$V$_2\hkl(530) (\SI{0.1537}{\nano\metre})$.
Given the phase diagrams for Ga-V and Ru-V \cite{okamoto_bapd}, we attribute this phase to a Ga-V binary such as Ga$_5$V$_2$ or Ga$_{41}$V$_8$, both of which are peritectics with relatively low decomposition temperatures. 
Based on the substantial increase in $S$, reordering of Ru from the Heusler $4d$ site occurs forming a highly disordered cubic alloy, with an A2 structure. 

There is a moderate \hl{CoPt$_3$} \hkl(111) and \hkl(200) texture before annealing. After annealing, only a strong \hl{CoPt$_3$}\hkl(200) peak is present, and magnetic anisotropy is also in-plane. 
\hl{In addition, the average roughness for Co increases moderately, while the roughness of the Pt layers approximately doubles in both cases.}

\subsection{Titanium Nitride}\label{subsec:TiN}
When using a TiN spacer, we observe a collapse in the tetragonality of MRG while $L_c$ increases and \hl{$\epsilon$ decreases}, indicating that defects are being removed. This is accompanied by a loss of coercivity and squareness in the magnetic hysteresis.
TiN is a well known diffusion barrier, and defects introduced by diffusion are not considered as the cause.
We also see that there is no direct correlation with Ru-Mn disorder, as there are three cases with small increase in $S$ contrasted by the \SI{1.4}{\nano\metre} TiN grown at high temperature which has a lower $S$ ratio. \hl{However, there is a correlation between strain $\epsilon$ and chemical ordering $S$.}

This means that defects such as stacking faults and grain boundaries induce strain on the crystal structure, which contributes strongly to the magneto-crystalline anisotropy of the film and has a more prominent effect than Ru-Mn anti-sites.
The effect is independent of growth temperature.

From the Co/Pt superlattice, we observe a very weak Pt\hkl(111) peak, which is distinct from the \hl{CoPt$_3$}\hkl(111), along with a strong \hl{CoPt$_3$}\hkl(200). The high intensity, as well as the Laue fringes observed around the peak indicates that the Pt\hkl(111) is due to the thick Pt capping layer, caused by relaxation at a critical thickness. The close lattice matching of TiN\hkl(100) to MRG\hkl(110) means the TiN\hkl(001) out-of-plane texture is expected, which helps to induce a \hl{CoPt$_3$}\hkl(200) \hl{phase} in the Co/Pt \cite{hongyu_2015}. However, PMA is lost after annealing.
\hl{Average roughness of Co and Pt increase moderately in all cases. Pt layers begin with a higher roughness in the case of high temperature grown TiN spacer layers. By comparison, the roughness after annealing is equal or greater when using room temperature grown TiN. This indicates that a high temperature grown TiN seed layer is less suitable, but is also more stable.}

\subsection{Hafnium Oxide}\label{subsec:HfO$_x$}
As HfO$_x$ is formed by natural oxidation of a Hf metal layer, there is a gradient of oxygen content that depends on the film thickness.
MRG performs poorly with a $\SI{1.4}{\nano\metre}$ thick HfO$_x$ spacer resulting in near total loss of perpendicular anisotropy upon annealing, whereas with a $\SI{2}{\nano\metre}$ spacer it maintains both coercivity and squareness with reductions of \SI{11}{\percent} and \SI{14}{\percent} respectively, \hl{despite a significant increase in strain ($\sim {\SI{25}{\percent}}$)}. 
This is correlated with a lesser reduction in $c$-spacing for the $\SI{2}{\nano\metre}$ spacer stack.
Similarly to the case of TiN, the reduction in $c$ comes despite an improvement in the coherence length. 

If the $\SI{1.4}{\nano\metre}$ HfO$_x$ is fully oxidised ($x \sim 2$), then diffusion into MRG, either across the barrier or from the barrier itself, will be reduced. This will result in oxidation of MRG at the interface, however. The effect is then similar to that using TiN.
\hl{As the HfO$_x$ films were treated for equal time, the} thicker HfO$_x$ is not fully oxidised due to passivation, which leads to some interdiffusion.

There is no observable texture in the Co/Pt superlattice before annealing. Weak \hl{CoPt$_3$}\hkl(111) and \hkl(200) texture is observed after annealing, but PMA is lost.
\hl{As with the TiN layers, we see a moderate increase in Co and Pt roughness, with a higher as-deposited roughness.}

\subsection{Hafnium}\label{subsec:Hf}
Hf spacers work well, with the $\SI{1.4}{\nano\metre}$ spacer showing only a small change in all properties.
In contrast to other materials, the $\SI{2}{\nano\metre}$ Hf spacer increases the $\delta_c$ of the MRG unit cell. Furthermore, it raises $T_{\text{comp}}$ to a value slightly above RT,
which causes $H_c$ to increase significantly due to its divergent nature close to $T_{\text{comp}}$.
The increase in $c$ is accompanied by a decrease in $L_c$ \hl{and an increase in $\epsilon$}, indicating an increase of line and planar defects.
According to the binary phase diagrams \cite{okamoto_bapd}, Hf absorption into both Mn and Ga is limited to $\sim \SI{1}{at\ldotp\thinspace\percent}$ while maintaining crystal structure. 
This limited solubility accounts for the reduced effects seen in the HfO$_x$ $\SI{2}{\nano\metre}$ and Hf $\SI{1.4}{\nano\metre}$ based heterostructures, where the counter-diffusion into the Hf layer limits the available soluble material.

We see different effects on the Co/Pt layer depending on the thickness of the Hf spacer, with the $\SI{1.4}{\nano\metre}$ thick spacer inducing no discernable texture in the superlattice. After annealing there is a weak \hl{CoPt$_3$}\hkl(111) texture. The $\SI{2}{\nano\metre}$ thick spacer promotes a \hl{CoPt$_3$}\hkl(111) orientation which is maintained after annealing, as well. In neither case is there any PMA after annealing, although there is a clear preferred orientation. 
\hl{The roughness of Co layers increases moderately. Annealing causes a substantial increase in roughness of Pt layers, to {$\sim$} {\SI{1}{\nano\metre}}, with the {\SI{2}{\nano\metre}} Hf spacer giving a higher initial roughness.}

\subsection{Molybdenum}\label{subsec:Mo}
Mo spacers have similar properties to Hf spacers, and the phase diagrams show a greater affinity of Mn for Mo ($\SI{4}{at\ldotp\thinspace\percent}$) and Ga ($\SI{16}{at\ldotp\thinspace\percent}$) Counter-diffusion is limited to Mn, as the Mo does not absorb much Ga at \SI{350}{\degreeCelsius}.
The increase in $S$ for the \SI{1.4}{\nano\metre} Mo heterostructure, indicates reordering of the Ru within MRG which should result in a decrease in magnetic properties. 
However, the squareness is still within acceptable limits and the coercivity increases, indicating that $T_{\text{comp}}$ has increased despite a lower $c$ parameter. This is corroborated by the change in sign of the MRG switching direction in EHE, showing that $T_{\text{comp}}$ is now above RT.

This effect is less prominent with the \SI{2}{\nano\metre} Mo heterostructure, but still indicates an increase in $T_{\text{comp}}$.
In this case $L_c$ \hl{and $\epsilon$ have both} increased without any change in $S$.
With more material to diffuse, the additional Mo \hl{inclusions maintain chemical ordering of the Ru $4d$ site.}

With the thin $\SI{1.4}{\nano\metre}$ spacer, a weak \hl{CoPt$_3$}\hkl(200) peak is present, which becomes stronger after annealing. By contrast, there is a clear \hl{CoPt$_3$}\hkl(220) peak present when using a $\SI{2}{\nano\metre}$ spacer layer, which remains after annealing. This is the only case we witnessed where magnetic anisotropy of the Co/Pt was still perpendicular to the film surface after annealing.
There is no clear Mo texture visible in XRD.
\hl{Co and Pt layers both have moderate increases of roughness, however the heterostructure with a {$\SI{2}{\nano\metre}$} spacer layer has higher initial and final roughness.}


\begin{table}[t]
\caption{\hl{Data for Co/Pt layers before and after annealing - XRD peaks visible are given by CoPt$_3${\hkl(111)} $\vartriangle$, CoPt$_3${\hkl(200)} $\circ$, CoPt$_3${\hkl(220)} $\square$, Pt{\hkl(111)} $\triangledown$. Filled shapes indicate a strong peak, underlined shapes indicate a weak peak. Average interfacial roughness of Co and Pt layers with the superlattice are given based on fitting of XRR data. A $\checkmark$ indicates that PMA was still present after annealing}}
\begin{tabularx}{255pt}{l X c c X c c }
\hline
 & \multicolumn{3}{c}{\textbf{As-deposited}} & \multicolumn{3}{c}{\textbf{Annealed}}\\
 & & \multicolumn{2}{c}{$\mathbf{R_a}$ (\si{\nano\metre})} &  & \multicolumn{2}{c}{$\mathbf{R_a}$ (\si{\nano\metre)}} \\
 \textbf{Spacer} & \textbf{XRD} & Co & Pt & \textbf{XRD}& Co & Pt\\
\hline 
V \SI{1.4}{\nano\metre}         & $\vartriangle \circ$     &0.44     &0.39    & $\bullet$     & 0.51  & 0.80  \\
V \SI{2}{\nano\metre}           & $\vartriangle \circ$     &0.43        &0.42       & $\bullet$     &0.52      &0.79      \\
TiN HT \SI{1.4}{\nano\metre}    & $\circ$ \underline{$\triangledown$} &0.40     &0.70       & $\bullet$ \underline{$\triangledown$} &0.45   &0.80   \\
TiN HT \SI{2}{\nano\metre}      & $\circ$ \underline{$\triangledown$} &0.43     &0.65       & $\bullet$ \underline{$\triangledown$} &0.50   &0.79   \\
TiN RT \SI{1.4}{\nano\metre}    & $\circ$ \underline{$\triangledown$} &0.41     &0.54       & $\bullet$ \underline{$\triangledown$} &0.47   &0.91   \\
TiN RT \SI{2}{\nano\metre}      & $\circ$ \underline{$\triangledown$} &0.44     &0.57       & $\bullet$ \underline{$\triangledown$} &0.50   &0.75   \\
HfO$_x$ \SI{1.4}{\nano\metre}   &   &0.44   &0.77   & \underline{$\vartriangle$} \underline{$\circ$} &0.49   &0.90    \\
HfO$_x$ \SI{2}{\nano\metre}     &   &0.44   &0.80   & \underline{$\vartriangle$} \underline{$\circ$} &0.55   &0.91     \\
Hf \SI{1.4}{\nano\metre}        &   &0.46   &0.53   & \underline{$\vartriangle$}    &0.50       &0.99         \\
Hf \SI{2}{\nano\metre}          & $\vartriangle$    &0.44   &0.73    & $\vartriangle$   &0.50   &1.00  \\
Mo \SI{1.4}{\nano\metre}        & \underline{$\vartriangle$}    &0.43   &0.43   & $\vartriangle$   &0.48   &0.50     \\
Mo \SI{2}{\nano\metre}  $\checkmark$        & $\square$     &0.39   &0.59   & $\blacksquare$    &0.45   &0.70     \\
\end{tabularx}
\label{table:copt}
\end{table}

\section{Discussion}\label{sec:disc}
Strain plays a significant role in the properties of MRG thin films, and crystalline point defects have previously been discussed as a source of electronic doping that maintains strain within the lattice \cite{zic_2016}.
As demonstrated by use of a TiN spacer, annealing of MRG removes line and planar-type crystalline defects, which relaxes the lattice. 
\hl{These types of defects have previously been identified in MRG from HRTEM} \cite{zic_2016,teichert2021}.
This results in a \SIrange{60}{70}{\percent} loss of tetragonal distortion of the unit cell (where a \SI{100}{\percent} reduction would indicate a fully cubic cell).
It is clear here that this is responsible for the reduction of coercivity and squareness in magnetic hysteresis due to the reorientation
of the magneto-crystalline anisotropy into the plane.  This effect is much stronger than could be attributable to Ru-based anti-sites, so it is not possible here to determine \hl{their contribution}.
\hl{Where diffusion is discounted, there is a correlation between strain and chemical ordering within MRG, consistent with }\cite{zic_2016}.

The diffusion of different materials has varying effect on MRG, \hl{for example} with V causing significant disorder of the crystal structure, ultimately resulting in the formation of additional intermetallic phases.
\hl{The choice of a suitable spacer material is based on a requirement to maintain magnetic properties of MRG after processing.}
Hf and Mo are strong candidates for future applications, 
\hl{as in both cases, the magnetic hysteresis is reliable with a change in squareness {\SI{<10}{\percent}} from the as-deposited state. This is especially important in magnetoresistive devices which rely upon the relative orientation of magnetic moments between the free and reference layers.}
$T_{\text{comp}}$ is sensitive to the number of valence electrons within MRG, and increases linearly with $n_v$ \cite{siewierska2021}. Both Hf and Mo appear to donate valence electrons to the film, as indicated by the increase in $T_{\text{comp}}$.
Based on Slater-Pauling rules, both Hf and Mo would donate $\sim 0.2$ electrons per unit cell of MRG for the considered solubility.
This contextualises the same effect seen in our previous work using thin interlayers \cite{borisov2017,titova2019}.
\hl{The results for these materials indicate a complex interplay between strain and film defects when we consider incorporation of additional valence electrons into the unit cell, that results in retention of perpendicular anisotropy.}

The strongly bonded compounds TiN and HfO$_x$ do not diffuse. However, the heterostructure using the thicker, partially oxidised HfO$_x$ spacer layer more resembles the structures using a Hf spacer layer in behaviour due to diffusion of metallic species closer to the MRG interface where the layer is not fully oxidised.

Typically we see either \hl{CoPt$_3$}\hkl(200) or \hl{CoPt$_3$}\hkl(111) texture in the free layer, however in the case of the \SI{2}{\nano\metre} Mo heterostructure there appears a \hl{CoPt$_3$}\hkl(220) peak. It is of interest that of all the structures, this is the only one that maintained perpendicular anisotropy in the Co/Pt after annealing.
We can surmise that a mixture of weak \hkl(111) and \hkl(200) indicates an amorphous interface between spacer and the first Co layer, which is then crystallised and induces a preferential texture on the superlattice during annealing.

Annealing results in an increase in interfacial roughness within the superlattice. 
\hl{PMA after annealing is independent on the degree of roughness measured, and is instead correlated with the appearance of CoPt$_3$ phases which are the result of intermixing. The superlattice which forms a CoPt$_3$}\hkl(220) phase does not loose perpendicular anisotropy, which we attribute here to the additional strain on the \hl{remaining Co/Pt interfaces} induced by the crystal texture.
As Mo is an element of comparable atomic weight to Pt, the additional orbital contribution at the interface between spacer and superlattice likely contributes to the anisotropy rather than any strain induced by epitaxial growth, considering that the Mo has no visible texture itself.
\hl{PMA after annealing when using a {$\SI{2}{\nano\metre}$} Mo spacer, when there is none present using a {$\SI{1.4}{\nano\metre}$} Mo spacer, indicates that counter diffusion of MRG has an effect on the interfacial relationship between the spacer and the initial Co layer of the superlattice.}

\section{Conclusions}\label{sec:conclusions}

\hl{This} study has enabled us to characterise the effects of \hl{various} spacer layers on the structural and magnetic properties of both the MRG and the Co/Pt superlattice, which will be useful for developing perpendicular GMR and TMR structures that can withstand annealing at \SI{350}{\degreeCelsius}. The PMA of MRG is more robust, since it depends on strain imposed by the MgO substrate. That of the Co/Pt multilayer is only maintained when a \hl{CoPt$_3${\hkl(220)} phase is developed by intermixing}, with the help of Mo.
\hl{Based on the effect of annealing MRG films capped with a TiN spacer, line and planar defects help to maintain the substrate-induced strain and the tetragonal distortion of the MRG crystal structure, which the perpendicular magnetic anisotropy is dependent on.}
Annealing causes these defects to be removed, which results in the partial collapse of tetragonal distortion and therefore magnetic properties dependent on the perpendicular anisotropy.
There was no identifiable link between these properties and the presence of Ru-based anti-site disorder.
\hl{The effect of strain on anisotropy is counteracted by the addition of valence electrons in the MRG unit cell}

In Co/Pt multilayers, interface roughness is the main antagonist of \hl{PMA due to the reduction of interfacial anisotropy. Intermixing results in the formation of a CoPt$_3$ phase, even before annealing. However, the development of CoPt$_3$} with \hkl(220) texture retains perpendicular anisotropy after a \SI{350}{\degreeCelsius} anneal.
A Mo underlayer is proposed here as a viable solution.

Hf and Mo \hl{will} be useful as thin protective layers in future active devices \hl{based on MRG}, as they \hl{are able to offset changes in strain and defect density to maintain the PMA of the MRG layer}.
The solubility of both materials within MRG provides additional valence electrons, modifying the Fermi level position, and therefore $T_{\text{comp}}$ \hl{as well as the overall magnetic properties}. This is in contrast to V, which prefers to form intermetallic phases instead. Considering the enhanced high-temperature stability of the crystal, doping of MRG with these or similar materials, such as Zr, should be investigated. By tuning the Ru/dopant composition, desired magnetic properties should be easily achievable.

\section{Acknowledgements}
This work was partly funded by Science Foundation Ireland under grant 12/RC/2278 (AMBER), and by the European Commission under Grant Agreement No. DLV-737038 (TRANSPIRE). The work of K. S. was supported by the Irish Research Council under Grant GOIPG/2016/308.

\bibliographystyle{elsarticle-num}
\bibliography{MRG_CoPt}
\end{document}